\documentclass[12pt, prd, reprint, amsmath, amssymb, aps, superscriptaddress,preprintnumbers]{revtex4-1}

\usepackage{appendix}
\usepackage{color}
\usepackage[dvipsnames]{xcolor}
\usepackage{graphicx}
\usepackage{siunitx}
\usepackage[citecolor=blue, linkcolor=blue, urlcolor=blue, colorlinks=true]{hyperref}

\newcommand{\be}{\begin{equation}}
\newcommand{\ee}{\end{equation}}
\newcommand{\bea}{\begin{eqnarray}}
\newcommand{\eea}{\end{eqnarray}}
\newcommand{\beq}{\begin{equation}}
\newcommand{\eeq}{\end{equation}}
\newcommand{\beqa}{\begin{eqnarray}}
\newcommand{\eeqa}{\end{eqnarray}}

\def\lsim{\mathrel{\rlap{\lower4pt\hbox{\hskip1pt$\sim$}}
    \raise1pt\hbox{$<$}}}         
\def\gsim{\mathrel{\rlap{\lower4pt\hbox{\hskip1pt$\sim$}}
    \raise1pt\hbox{$>$}}}         

\newcommand{\unit}[1]{\,\ensuremath{\textrm{#1}}}

\newcommand{\tev}{\unit{TeV}}

\newcommand{\ifb}{\unit{fb}^{-1}}

\newcommand{\cp}{$\mathcal{CP}$}

\begin{document}

\title{Implications of the upper bound on $\boldsymbol{h\to\mu^+\mu^-}$\\
on the baryon asymmetry of the Universe}

\author{Elina Fuchs}
\email{elina.fuchs@weizmann.ac.il}
\affiliation{Department of Particle Physics and Astrophysics, Weizmann Institute of Science, Rehovot, Israel 7610001}
\affiliation{Fermilab, Theory Department, Batavia, IL 60510, USA}
\affiliation{University of Chicago, Department of Physics, Chicago, IL 60637, USA}

\author{Marta Losada}
\email{marta.losada@nyu.edu}
\affiliation{New York University Abu Dhabi, PO Box 129188, Saadiyat Island, Abu Dhabi, United Arab Emirates}

\author{Yosef Nir}
\email{yosef.nir@weizmann.ac.il}
\affiliation{Department of Particle Physics and Astrophysics, Weizmann Institute of Science, Rehovot, Israel 7610001}

\author{Yehonatan Viernik}
\email{yehonatan.viernik@weizmann.ac.il}
\affiliation{Department of Particle Physics and Astrophysics, Weizmann Institute of Science, Rehovot, Israel 7610001}

\preprint{FERMILAB-PUB-19-622-T, EFI-19-13}

\begin{abstract}
\noindent
The upper bounds from the ATLAS and CMS experiments on the decay rate of the Higgs boson to two muons provide the strongest constraint on an imaginary part of the muon Yukawa coupling. This bound is more than an order of magnitude stronger than bounds from \cp{}-violating observables, specifically the electric dipole moment of the electron. It excludes a scenario -- which had been viable prior to these measurements -- that a complex muon Yukawa coupling is the dominant source of the baryon asymmetry. Even with this bound, the muon source can still contribute ${\cal O}(16\%)$ of the asymmetry.
\end{abstract}

\maketitle

\section{Introduction}
Following the discovery of the Higgs boson by the ATLAS and CMS experiments \cite{Aad:2012tfa,Chatrchyan:2012xdj}, an intensive program to study its properties has been pursued. This experimental program is relevant to many open questions in particle physics and in cosmology \cite{Heinemann:2019trx}. One of the most intriguing questions relates to the fact that the Kobayashi-Maskawa phase of the Standard Model (SM) fails to account for the baryon asymmetry of the Universe by many orders of magnitude~\cite{Gavela:1993ts,Huet:1994jb}, and thus additional sources of \cp{} violation must exist in Nature. In this work we show that searches at the LHC for the Higgs decay to two muons, $h\to\mu^+\mu^-$, give a definite answer to the question of whether a complex Yukawa coupling of the muon can provide the \cp{} violation necessary for baryogenesis.

There are several reasons why we focus our discussion on the muon. First, it is a rather unique case where ATLAS and CMS measurements give a definite answer to an open question in particle cosmology. Second, while the question of whether complex Yukawa couplings of third generation fermions can account for the CP asymmetry that is necessary for baryogenesis was studied before, we are not aware of previous studies of whether a second generation fermion can provide the dominant source. 
Finally, the measurement of the rate of $h\to\mu^+\mu^-$ is
flavor-specific, while the contributions of various fermions to the baryon asymmetry are close to
being additive, thus our final conclusions would not be affected by taking into account
additional complex Yukawa couplings.

The possibility that the baryon asymmetry is generated by electroweak baryogenesis requires the SM to be extended in two ways: The scalar potential has to be modified in such a way that the electroweak phase transition is strongly first order, and new sources of \cp{} violation must be introduced. The former aspect has been intensively discussed in the literature (for reviews see e.g.~Refs.~\cite{Cline:2006ts,Morrissey:2012db}), and we do not elaborate on it here. For the latter, we ask whether the {\it dominant} source of \cp{} violation can be a complex Yukawa coupling of the muon, coming from a dimension-six term in the SM effective field theory (SMEFT). Thus, similar to the studies in Ref. \cite{deVries:2017ncy,deVries:2018tgs}, we assume the following:
\begin{itemize}
\item Whatever the extension of the SM that modifies the nature of the electroweak phase transition, it does not affect in a significant way the other aspects of the baryon asymmetry, such as the rates of \cp{}-conserving and \cp{}-violating fermion processes in the transport equations. Examples of such extensions include the addition of a real scalar field \cite{Huang:2016cjm,Craig:2013xia,Curtin:2014jma} or the addition of dimension-six $(H^\dagger H)^3$ term \cite{Grojean:2004xa}.
\item The new degrees of freedom that are relevant to \cp{} violation are heavy enough that their effects on electroweak baryogenesis and on the $h\to\mu^+\mu^-$ decay can be represented by the SMEFT.
\end{itemize}

The plan of this paper is as follows. In Section \ref{sec:framework} we introduce our theoretical framework and notations, and derive the effective muon Yukawa coupling in this framework. The contributions of the complex Yukawa coupling to the rate of Higgs decay to two muons, $\Gamma(h\to\mu^+\mu^-)$, to the electric dipole moment (EDM) of the electron, $d_e$, and to the baryon asymmetry of the Universe, $Y_B$, are discussed in Sections \ref{sec:mumumu}, \ref{sec:de} and \ref{sec:yb}, respectively. In Section \ref{sec:interplay} we analyze the interplay between these three observables, and reach our conclusions.

\section{The framework}
\label{sec:framework}
Within the SMEFT with terms up to dimension (dim) six, the muon mass and the muon Yukawa coupling arise from the following terms:
\beq\label{eq:lagmu}
{\cal L}_{\rm Yuk}^\mu= y_\mu\overline{L_{\mu}}\mu_RH+\frac{1}{\Lambda^2}(X_R^\mu+iX_I^\mu)|H|^2\overline{L_{\mu}}\mu_RH+{\rm h.c.},
\eeq
where $L_{\mu}=(\nu_\mu\ \mu_L^-)^T$ is the left-handed muon doublet, $\mu_R$ is the right-handed muon singlet, $H=(0\ \frac{v+h}{\sqrt2})^T$ is the Higgs doublet, and $\Lambda$ is the scale of new physics. The couplings $y_\mu$, $X_R^\mu$ and $X_I^\mu$ are dimensionless and, without loss of generality, we take $y_\mu$ to be real. From here on, we omit the flavor index $\mu$ from $y,X_R,X_I$.

We are interested in obtaining the muon mass $m_\mu$ and $h\mu\mu$ Yukawa coupling $\lambda_\mu$:
\beq\label{eq:yukd46}
{\cal L}^\mu\supset m_\mu\overline{\mu_L}\mu_R
+\frac{\lambda_\mu}{\sqrt2}\overline{\mu_L}\mu_R h+{\rm h.c.}\,.
\eeq
It is convenient to define
\begin{align}
\label{eq:TRTIdef}
T_R\equiv\frac{v^2}{2\Lambda^2}\frac{X_R}{y},~~~
T_I\equiv\frac{v^2}{2\Lambda^2}\frac{X_I}{y}.
\end{align}
We obtain:
\beq\label{eq:mlambdad46}
m_\mu=\frac{y v}{\sqrt2}\left(1+T_R+iT_I\right),\ \
\lambda_\mu=\frac{y}{\sqrt2}\left(1+3T_R+3iT_I\right).
\eeq
Transforming to a basis where the muon mass is real, we have
\beqa\label{eq:massyuk}
m_\mu&=&\frac{y v}{\sqrt2}\sqrt{(1+T_R)^2+T_I^2}\,,\\
\lambda_\mu&=&\frac{y}{\sqrt2}\frac{1+4T_R+3T_R^2+3T_I^2+2iT_I}{\sqrt{(1+T_R)^2+T_I^2}}\,
\eeqa
In this basis,
\beq
{\cal I}m(\lambda_\mu)=\frac{v}{m_\mu}y^2T_I.
\eeq
An equivalent statement, valid in any basis, is that ${\cal I}m(m_\mu^*\lambda_\mu)=vy^2T_I$.

As we will see below, the various constraints allow $|T_{R,I}|\not\ll1$, and thus the terms of ${\cal O}(T_{R,I}^2)$ can be non-negligible. These terms are of order $v^4/\Lambda^4$, yet taking them into consideration and not the 
dim-8 
terms of the SMEFT is a consistent procedure. This is due to the smallness of the 
dim-4 
Yukawa coupling of the muon: at order $v^4/\Lambda^4$, the contribution from the product of the dim-4 and dim-8 terms will be suppressed by $y_\mu$ compared to the contribution from the dim-6 term squared.

The modification of the SM relation between the Yukawa coupling and the mass, $\lambda_\mu\neq m_\mu/v$, and, in particular, the generation of an imaginary part, ${\cal I}m(\lambda_\mu/m_\mu)\neq0$, entail interesting consequences:
\begin{itemize}\item The decay rate of the Higgs boson to two muons, $\Gamma(h\to\mu^+\mu^-)$, is modified;
\item The muon Yukawa coupling contributes to the EDM of the electron $d_e$;
\item The muon Yukawa coupling contributes to the baryon asymmetry $Y_B$.
\end{itemize}
These observables will be discussed in the next three sections.

\section{The $ \textit{\textbf{h}} $ \boldmath{$\to \mu^+\mu^-$} decay}
\label{sec:mumumu}
The ATLAS and CMS experiments report their measurements of $pp\to h\to f\bar f$ via
\beq \label{eq:muff}
\mu_{f\bar f}\equiv\frac{\sigma_i(pp\to h){\rm BR}(h\to f\bar f)}{[\sigma_i(pp\to h){\rm BR}(h\to f\bar f)]_{\rm SM}},
\eeq
where $\sigma_i(pp\to h)$ denotes the cross section of a specific Higgs production mode $i$, such as gluon-gluon fusion (ggF) or vector-boson fusion (VBF). If the contribution of the 
dim-6 
terms modifies neither the Higgs production cross section, nor the total Higgs width in a significant way, as is the case for ${\cal O}(1)$ (or smaller) modification of $\lambda_\mu$, then Eq.~(\ref{eq:muff}) simplifies to
\beq
\mu_{\mu^+\mu^-}=
\frac{\Gamma(h\to \mu^+\mu^-)}{[\Gamma(h\to \mu^+\mu^-)]_{\rm SM}}.
\eeq
Using Eq.~(\ref{eq:mlambdad46}), we obtain
\beq\label{eq:mumumuth}
\mu_{\mu^+\mu^-}=\frac{(1+3T_R)^2+9T_I^2}{(1+T_R)^2+T_I^2}.
\eeq
Taking into account that $y^{\rm SM}_f=\sqrt2 m_f/v$, we can write
\begin{align}
\label{eq:mumy}
1&=(y/y^{\rm SM})^2\left[(1+T_R)^2+T_I^2\right],\\
\mu_{\mu^+\mu^-}&=(y/y^{\rm SM})^2\left[(1+3T_R)^2+9T_I^2\right]\,.
\end{align}

An upper bound $\mu_{\mu^+\mu^-}\leq \mu^{\rm max}$, yields then the following bounds on $|T_I|$, on $y|T_I|$, and on $y^2|T_I|$,
\begin{align}
|T_I| &\leq\frac{2\sqrt{\mu^{\rm max}}}{9-\mu^{\rm max}}\,,\label{eq:timax}\\
(y/y^{\rm SM})|T_I| &\leq{\rm min}\left[\frac{\sqrt{\mu^{\rm max}}}{3},1\right],\label{eq:ytimax}\\
(y/y^{\rm SM})^2|T_I| &\leq\frac{\sqrt{\mu^{\rm max}}}{2}\,.\label{eq:yytimax}
\end{align}
We note that bounds of the form (\ref{eq:timax}) -- (\ref{eq:yytimax}) apply to any fermion, as long as the Higgs production rate and total width are not significantly modified. Such a case will be discussed in Ref.~\cite{Fuchs:2020uoc}.

In fact, at present there are only upper bounds on $\mu_{\mu^+\mu^-}$
by CMS combining the $\sqrt{s}=7$ and $8\tev$ data sets with $35.9\ifb$ at $13\tev$~\cite{Sirunyan:2018hbu}
and by ATLAS with the full data set of $139\ifb$ at $13\tev$~\cite{ATLAS:2019ain}:
%
\begin{align}
\mu_{\mu^+\mu^-}^{\rm CMS}  &<2.9\ {\rm at}\ 95\%\ {\rm C.L.}\,,\label{eq:mumumuCMS}\\
\mu_{\mu^+\mu^-}^{\rm ATLAS}&<1.7\ {\rm at}\ 95\%\ {\rm C.L.}\,.\label{eq:mumumuATLAS}
\end{align}
A bound on the signal strength of Eq.~(\ref{eq:mumumuth})
translates into an allowed region within a circle in the $T_R-T_I$ plane, centered around $(T_R, T_I)=\left(-1+\frac{6}{9-\mu},0\right)$ with a radius of $\frac{2\sqrt{\mu}}{9-\mu}$.
The bound from Eq.~(\ref{eq:mumumuATLAS}) is plotted in Fig. \ref{fig:collTRTI_mu}. It provides the allowed range for $T_R$:
\beq\label{eq:mutr}
-0.5\lsim T_R\lsim0.2,
\eeq
and the following upper bounds on \cp{} violation from the complex Yukawa coupling of the muon:
\begin{align}
|T_I| &\leq0.36\,,\label{eq:muti}\\
(y/y^{\rm SM})|T_I| &\lsim0.44\,,\label{eq:muyti}\\
(y/y^{\rm SM})^2|T_I| &\lsim0.65\,.\label{eq:muy2ti}
\end{align}
%

\section{The EDM of the electron}
\label{sec:de}
The dimension-six term in the Lagrangian contributes to the electric dipole moment of the electron \cite{Panico:2018hal}:
\beq\label{eq:demupar}
\frac{d_e^{(\mu)}}{e}\simeq-4 Q_\mu^2\frac{e^2}{(16\pi^2)^2}\frac{m_em_\mu}{m_h^2}\frac{v}{\Lambda^2}
X_I^{\mu}\left(\frac{\pi^2}{3}+\ln^2\frac{m_\mu^2}{m_h^2}\right),
\eeq
where $Q_\mu=-1$ is the electromagnetic charge of the muon, and the equation is written in the basis where $m_\mu$ is real. Eq.~(\ref{eq:demupar}) translates into the following numerical estimate:
\beq\label{eq:demunum}
d_e^{(\mu)}\simeq-1.0\times10^{-30}\ (y/y^{\rm SM})^2T_I\ e\ {\rm cm}.
\eeq
Given the upper bound of Eq.~(\ref{eq:muy2ti}), we obtain an upper bound on the contribution to $d_e$ from a complex muon Yukawa coupling:
\beq
|d_e^{(\mu)}|\leq 6.5\times10^{-31}\ e\ {\rm cm}.
\eeq

The ACME collaboration provided an upper bound on $|d_e|$ at 90\% CL \cite{Andreev:2018ayy}:
\beq
|d_e^{\rm max}|=1.1\times10^{-29}\ e\ {\rm cm}.
\eeq
We learn that the bound on a \cp{} violating muon Yukawa coupling from the measurement of the \cp{} conserving observable $\mu_{\mu^+\mu^-}$ is much stronger than the bound from the \cp{} violating observable $d_e$. To compete with the LHC current bound, the $d_e$ sensitivity has to improve by a factor of ${\cal O}(15)$.
For a comparison of EDM and LHC constraints on real and imaginary parts of Yukawa couplings of third-generation fermions see also Refs.~\cite{Brod:2013cka,Fuchs:2020uoc}.

\section{The Baryon Asymmetry}
\label{sec:yb}
We calculate the baryon asymmetry using the Closed Time Path formalism, similar to Ref.~\cite{deVries:2018tgs}. The details of our calculation and the innovations it introduces will be described in a more detailed report~\cite{flnv}, where we will also present the contributions from other modified Yukawa couplings and their combinations.

The process whereby the baryon asymmetry is generated by the complex Yukawa coupling of the muon can be summarized as follows. During the electroweak phase transition, the Yukawa interaction of the muon across the expanding wall produces a \cp{} asymmetry. While relaxation processes wash out the asymmetry in the broken phase, part of the asymmetry diffuses into the symmetric phase. 
For the muon, the washout processes are less efficient than for quarks or for tau-leptons,
which is helpful in overcoming the suppression of the asymmetry due to the smallness of the muon Yukawa coupling. Weak sphaleron interactions act on the net chiral density that has diffused into the symmetric phase, while strong sphaleron interactions, which would wash out asymmetries in the quark sector, do not act on the lepton sector. Finally, the bubble wall catches up with the region of net asymmetry, and freezes in the resulting baryon asymmetry in the broken phase.

The baryon asymmetry is proportional to the muon source $Y_B\propto S_\mu$, with $S_\mu\propto y^2T_I$. The $T_R$ dependence is mild and enters only at second order in $T_{R,I}$. Explicitly, the relaxation rate $\Gamma_M$ and the Yukawa rate $\Gamma_Y$ are modified in the presence of the dimension-six term as follows:
\begin{align}
\Gamma_M &\to \left[ \frac{(1+ r_{N0}^2T_R)^2 + (r_{N0}^2T_I)^2}{(1+T_R)^2 + T_I^2}\right]\,\Gamma_M\,, \\
\Gamma_Y &\to \left[ \frac{(1+3r_{N0}^2T_R)^2 + (3r_{N0}^2T_I)^2}{(1+T_R)^2 + T_I^2} \right]\,\Gamma_Y .
\end{align}
Here $r_{N0}\equiv v(T=T_N)/v(T=0)$, where $T_N$ is the nucleation temperature. For $T_R=0$, our calculation yields
\beq\label{eq:ybti}
Y_B^{(\mu)}=-2.1\times10^{-11}(y/y^{\rm SM})^2 T_I.
\eeq
Given the upper bound of Eq.~(\ref{eq:muy2ti}), we obtain an upper bound on the contribution to $Y_B$ from a complex muon Yukawa coupling:
\beq\label{eq:ybmumax}
|Y_B^{(\mu)}|\leq 1.4\times10^{-11}.
\eeq

The observed value of the baryon asymmetry was measured by PLANCK~\cite{Ade:2015xua,Tanabashi:2018oca} as
$\Omega_bh^2=0.02226(23)$ or, equivalently,
\beq
Y_B^{\rm obs}=(8.59\pm0.08)\times10^{-11}.
\eeq
Given the mild dependence on $T_R$, we conclude that a complex Yukawa coupling of the muon cannot account for the baryon asymmetry of the Universe. Yet, its contribution to the overall asymmetry created by different \cp{}-violating sources, could be non-negligible, of order $16\%$.

The calculation of the baryon asymmetry suffers from uncertainties in the rates of various washout processes and in the bubble wall parameters. We find, however, that these uncertainties have little effect on the contribution of the muon-Yukawa to the baryon asymmetry. For example, decreasing the relaxation rates by a factor of $O(10)$ raises the upper bound on $|Y_B^{(\mu)}|$ from order $16\%$ to order $20\%$ of the observed asymmetry. Increasing the relaxation rates makes the bound stronger. Other uncertainties have even smaller effect on our bound.

Due to the same scaling of $d_e$ and $Y_B$ as 
$(y/y^{\rm SM})^2 T_I$, see Eqs.~(\ref{eq:demunum}) and (\ref{eq:ybti}),
we obtain the following relation between the muon contributions to $d_e$ and to $Y_B$:
\beq
\frac{Y_B^{(\mu)}}{8.6\times10^{-11}}=\frac{d_e^{(\mu)}}{4.1\times10^{-30}\ e\ {\rm cm}}.
\eeq
This relation shows that the current bound on $d_e$ could not, by itself, exclude a scenario where a complex muon Yukawa coupling accounts for the baryon asymmetry.

\begin{figure}[h]
 \begin{center}
  \includegraphics[width=0.4\textwidth]{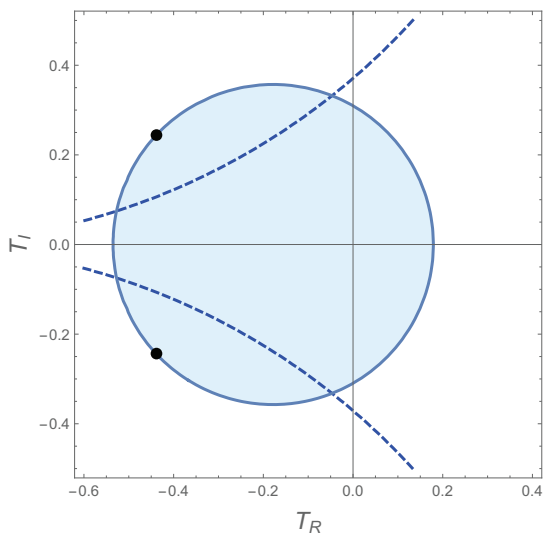}
  \caption{The blue region within the circle is the allowed region for $\mu_{\mu^+\mu^-}\leq1.7$~\cite{ATLAS:2019ain}. The two points on the circle correspond to maximal values of $|Y_B^{(\mu)}|_{\rm max}=1.4\times10^{-11}$ and of $|d_e^{(\mu)}|_{\rm max}=6.5\times10^{-31}\ e\ {\rm cm}$. The dashed lines correspond to $|Y_B^{(\mu)}|_{\rm max}/2$ and $|d_e^{(\mu)}|_{\rm max}/2$. 
  }
  \label{fig:collTRTI_mu}
 \end{center}
\end{figure}

\section{Discussion}
\label{sec:interplay}
Within the SM as a low-energy effective field theory, the leading modification to the SM Yukawa couplings and to their relation to the corresponding fermion masses comes from dimension-six terms. The contributions from these terms to the Yukawa couplings are suppressed by the ratio of scales, $v^2/\Lambda^2$. However, for small dimension-four Yukawa couplings, such contributions can be significant and even dominant.

In the case of the muon, which is the focus of this study, $y_\mu^{\rm SM}\sim6\times10^{-4}$, the contribution to $\lambda_\mu$ from the dimension-six term can be comparable or dominant for $\Lambda\lsim10$ TeV. If the dimension-six term dominates over the dimension-four, $(T_R^2+T_I^2)\gg1$, then $\mu_{\mu^+\mu^-}\simeq9$. The fact that the experimental upper bound is well below this value leads to a first important conclusion:
\begin{itemize}
\item The effective muon Yukawa coupling is not dominated by contributions from non-renormalizable terms.
\end{itemize}
The upper bounds on $|T_R|$ and $|T_I|$ in Eqs. (\ref{eq:mutr}) and (\ref{eq:muti}) can be translated into a lower bound on the scale $\Lambda$, by using Eq. (\ref{eq:TRTIdef}). We obtain 
\begin{equation}
\Lambda/\sqrt{X_{R,I}}\gsim 7\tev\,.
\end{equation}

The presence of dimension-six terms opens the door to a complex effective muon Yukawa coupling. Two \cp{}-violating observables can reveal the existence of this new source of \cp{} violation: The electric dipole moment of the electron and the baryon asymmetry of the Universe. Examining Eqs. (\ref{eq:demunum}) and (\ref{eq:ybti}) we learn that, interestingly, for $y\sim y^{\rm SM}$, the baryon asymmetry could have been generated by $\lambda_\mu$ as the dominant \cp{} violating source, while the $d_e$ bound is respected.

It is surprising then that the scenario of a complex muon Yukawa coupling generating the baryon asymmetry is unambiguously excluded by a measurement of a \cp{}-conserving observable, the Higgs decay rate into two muons, as can be seen by examining Eqs. (\ref{eq:muyti}) and (\ref{eq:ybti}). The strong upper bound on  $\mu_{\mu^+\mu^-}$ leads to a second important conclusion:
\begin{itemize}
\item The baryon asymmetry is not dominated by a complex muon Yukawa coupling.
\end{itemize}
Note that the modification of $\mu_{\mu^+\mu^-}$ depends quadratically on the \cp{} violating parameter, $[{\cal I}m(\lambda_\mu)]^2\lsim10^{-6}$. The fact that the leading constraint on this parameter comes from the ATLAS and CMS measurements shows the power of these experiments to probe very rare processes.

Yet, the muon contribution to the baryon asymmetry is not necessarily negligible.  We are led to a third conclusion:
\begin{itemize}
\item A complex $\lambda_\mu$ could account for as much as 16\% of $Y_B$, given current collider constraints.
\end{itemize}
In order to reach the observed baryon asymmetry, additional \cp{}-violating sources beyond the muon Yukawa coupling are needed -- in the considered dim-6 Yukawa framework thus also complex Yukawa couplings of other fermions.
If, in the future, $\mu_{\mu^+\mu^-}$ will be measured to be very close to 1, the bounds on the maximal contribution to the \cp{}-violating observables  will become stronger, but not by much: $|d_e^{(\mu)}/d_e^{\rm max}|\lsim0.05$ and $|Y_B^{(\mu)}/Y_B^{\rm obs}|\lsim0.12$. It may happen, however, that experiments will establish $\mu_{\mu^+\mu^-}<1$. 
While in this case the role of the muon for baryogenesis will be even smaller, such a measured deviation from the SM (likewise for establishing $\mu_{\mu^+\mu^-}>1$) will show that $h$ and $v$ are not aligned and/or Yukawa terms of dim$>4$ contribute. This will make it plausible that also third generation Yukawa couplings differ from their SM values and play a role in electroweak baryogenesis.

We conclude that the Higgs program at the LHC experiments leads to progress not only on open questions in particle physics, such as whether various Yukawa couplings are dominated by higher-dimensional terms, but also in particle cosmology, such as whether the baryon asymmetry is generated by complex Yukawa couplings. Specifically for the muon, the answers provided to both questions are negative. For other fermions, the answer to the latter question might be in the affirmative \cite{deVries:2018tgs}, as will be discussed in \cite{Fuchs:2020uoc}.


\acknowledgments{We are grateful to Jorinde van de Vis for very helpful discussions. We thank Daniel Aloni for interesting discussions during the early stages of this work.
ML would like to deeply thank the Weizmann Institute of Science for its hospitality during the completion of this work.
EF acknowledges the support by the Minerva Foundation.
YN is the Amos de-Shalit chair of theoretical physics, and is supported by grants from the Israel Science Foundation (grant number 394/16), the United States-Israel Binational Science Foundation (BSF), Jerusalem, Israel (grant number 2014230), and the I-CORE program of the Planning and Budgeting Committee and the Israel Science Foundation (grant number 1937/12).
}

\bibliography{muon}

\end{document}